\documentclass[]{article}
\usepackage{amsmath,booktabs}
\usepackage{epsfig,psfig}

\def\ie{{\em i.e.}, }
\begin{document}

\title{\large The topology  of the regulatory
interactions predicts the expression pattern of the segment polarity
genes in {\it Drosophila melanogaster}}
\author{\normalsize R\'eka Albert  and Hans G. Othmer\\ 
\normalsize Department of Mathematics, University of Minnesota,
Minneapolis, MN 55455\\
}
\date{}
\maketitle

\begin{abstract}
Expression of the {\it Drosophila} segment polarity genes is initiated
by a prepattern of pair-rule gene products and maintained by a network
of regulatory interactions throughout several stages of embryonic
development. Analysis of a model of gene interactions based on
differential equations showed that wild-type expression patterns of
these genes can be obtained for a wide range of kinetic parameters,
which suggests that the steady states are determined by the topology
of the network and the type of regulatory interactions between
components, not the detailed form of the rate laws. To investigate
this, we propose and analyze a Boolean model of this network which is
based on a binary ON/OFF representation of transcription and protein
levels, and in which the interactions are formulated as logical
functions. In this model the spatial and temporal patterns of gene
expression are determined by the topology of the network and whether
components are present or absent, rather than the absolute levels of
the mRNAs and proteins and the functional details of their
interactions. The model is able to reproduce the wild type gene
expression patterns, as well as the ectopic expression patterns
observed in over-expression experiments and various mutants. Furthermore, we
compute explicitly all steady states of the network and
identify the basin of attraction of each steady state. The
model gives important insights into the functioning of the segment
polarity gene network, such as the crucial role of the {\it wingless} and {\it
sloppy paired} genes, and the network's ability to correct errors in the 
prepattern.
\end{abstract}

\bigskip
\hrule
\bigskip

\section{Introduction}

Understanding how genetic information is translated into proteins in
the correct temporal sequence and spatial location to produce the
various cell types in an adult organism remains a major challenge in
developmental biology.  The sequencing of the genome of various
organisms provides the first step toward this understanding, but it is
equally important to understand the patterns of control involved in
gene regulation. Gene products often regulate their own production or
that of other proteins at any of a number of steps, and frequently the
result is a complex network of regulatory interactions.  Recent
experimental progress in dissecting the qualitative structure of many
signal transduction and gene control networks (see, for example,
Davidson {\it et al.}  2002) has produced a surge of interest in the
quantitative description of gene regulation. Broadly speaking, the
modeling approaches can be divided into two main groups. In the
`discrete-state' approach each network node (mRNA or protein) is
assumed to have a small number of discrete states and the regulatory
interactions between nodes are described by logical functions similar
to those used in programming. Typically time is also quantized, and
the network model that describes how gene products interact to
determine the state at the next time gives rise to a discrete
dynamical system (Bodnar 1997, Mendoza
{\it et al.} 1999, Bodnar \& Bradley 2001, S\'anchez \& Thieffry 2001,
Yuh {\it et al.} 2001). A more detailed level of description is used
in the `continuous-state' approach, in which the level of mRNAs,
proteins, and other components are assumed to be continuous functions
of time, and the evolution of these components within a cell is
modeled by differential equations with mass-action kinetics or other
rate laws for the production and decay of all components (Reinitz \&
Sharp 1995, von Dassow {\it et al.} 2000, Gursky {\it et al.}
2001). While it is widely thought that merely specifying the topology
of a control network (\ie the connections between nodes) places few
constraints on the dynamics of the network, our purpose here is to
demonstrate that in one well-characterized system, knowledge of the
interactions together with their signatures, by which we mean whether
an interaction is activating or inhibiting, is enough to reproduce the
main characteristics of the network dynamics.

\begin{figure}[htb]
\centerline{\psfig{figure=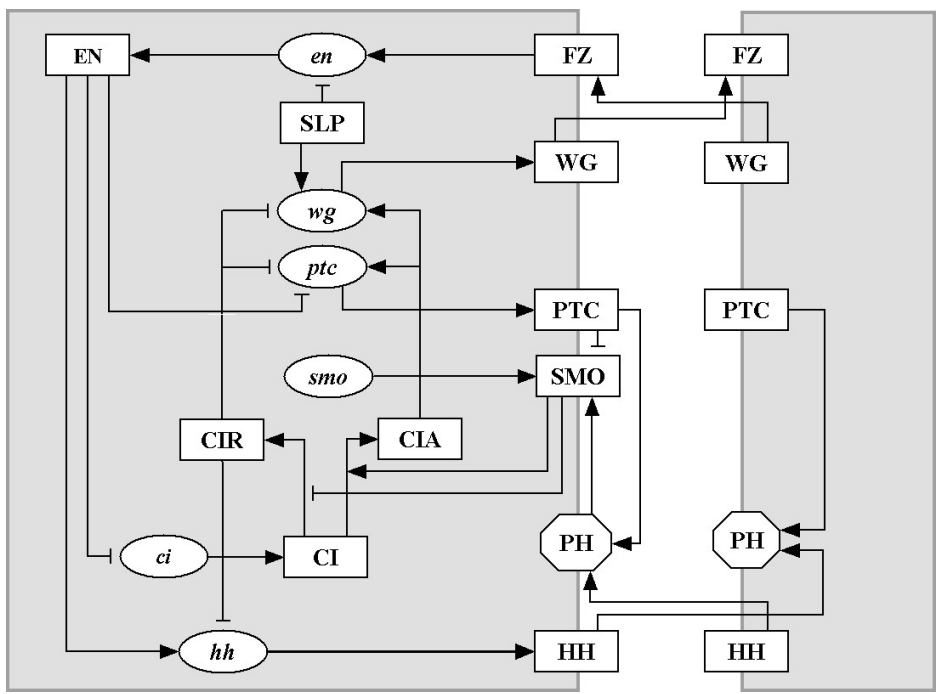,width=.55\textwidth,angle=-90}}
\caption{The network of interactions between the segment polarity 
genes. The shape of the nodes indicates whether the corresponding
substances are mRNAs (ellipses), proteins (rectangles) or protein
complexes (octagons). The edges of the network signify either biochemical 
reactions (e.g. translation) or regulatory interactions (e.g. transcriptional activation).
The edges are distinguished by their signatures, \ie whether they are activating
or inhibiting. Terminating arrows ($\rightarrow$) indicate translation,
post-translational modifications (in the case of CI), transcriptional
activation or the promotion of a post-translational modification
reaction (e.g., SMO determining the activation of CI). Terminating
segments ($\dashv$ ) indicate transcriptional inhibition or in the case of SMO,
the inhibition of the post-translational modification reaction
CI$\rightarrow$CIR.}
\label{fig_gene}
\end{figure}

The genes involved in embryonic pattern formation in the fruit fly
{\it Drosophila melanogaster}, as well as the majority of the
interactions between them, are known (for recent reviews see
Ingham-McMahon 2001, Sanson 2001, Hatini \& DiNardo 2001). As in other
arthropods, the body of the fruit fly is composed of segments, and
determination of the adult cell types in these segments is controlled
by about $40$ genes organized in a hierarchical cascade comprising the
gap genes, the pair-rule genes, and the segment polarity genes (Hooper
\& Scott 1992).  These genes are expressed in consecutive stages of
embryonic development in a spatial pattern that is successively more
precisely-defined, the genes at one step initiating or modulating the
expression of those involved in the next step of the cascade. While
most of these genes act only transiently, the segment polarity genes are 
expressed throughout the life of the fly. The segment polarity genes refine and maintain
their expression through the  network of intra- and intercellular
regulatory interactions shown in Figure \ref{fig_gene}. The stable
expression pattern of these genes (specifically the expression of {\it
wingless} and {\it engrailed}) defines and maintains the borders
between different parasegments (the embryonic counterparts of the
segments) and contributes to subsequent developmental processes,
including the formation of denticle patterns and of appendage
primordia (Hooper \& Scott 1992, Wolpert {\it et al.}  1998).
Homologs of the segment polarity genes have been identified in
vertebrates, including humans, which suggests strong evolutionary
conservation of these genes.

The segment polarity genes encode for the transcription factor
engrailed (EN), the cytosolic protein cubitus interruptus (CI), the
secreted proteins wingless (WG) and hedgehog (HH), and the
transmembrane receptor proteins patched (PTC) and smoothened (SMO)
involved in transduction of the HH signal. 

 The pair-rule gene {\it sloppy paired} ($slp$) is activated before
the segment polarity genes and expressed constitutively thereafter
(Grossniklaus 1992, Cadigan {\it et al.} 1994). {\it slp} encodes two
forkhead domain transcription factors with similar functions that
activate $wg$ transcription and repress $en$ transcription, and since
they are co-expressed we designate them both SLP. The $wg$ gene
encodes a glycoprotein that is secreted from the cells that synthesize
it (Hooper \& Scott 1992, Pfeiffer \& Vincent 1999), and can bind to
the Frizzled (FZ) receptor on neighboring cells. Binding of WG to the FZ
receptors on adjacent cells initiates a signaling cascade leading to
the transcription of {\it engrailed} ($en$) (Cadigan \& Nusse
1997). EN, the homeodomain-containing product of the {\it en} gene,
promotes the transcription of the {\it hedgehog} gene ({\it hh})
(Tabata {\it et al.} 1992). In addition to the homeodomain, EN
contains a separate repression domain (Han \& Manley 1993) that
affects the transcription of $ci$ (Eaton \& Kornberg 1990) and
possibly $ptc$ (Hidalgo \& Ingham 1990, Taylor {\it et al.} 1993). The
hedgehog protein (HH) is tethered to the cell membrane by a
cholesterol linkage that is severed by the dispatched protein (Ingham
2000), freeing it to bind to the HH receptor PTC on a neighboring cell
(Ingham \& McMahon 2001). The intracellular domain of PTC forms a
complex with smoothened (SMO) (van den Heuvel \& Ingham 1996) in which
SMO is inactivated by a post-translational conformation change (Ingham
1998).  Binding of HH to PTC removes the inhibition of SMO, and
activates a pathway that results in the modification of CI (Ingham
1998). CI contains at least three distinct domains: an $NH_2$ terminal
region characteristic of transcriptional repressors, a zinc finger
domain, and a $COOH$ region typical of activation domains in
transcription factors (Alexandre {\it et al.} 1996). The CI protein
can be converted into
one of two transcription factors, depending on the activity of
SMO. Several proteins have been implicated in this conversion,
including Fused, Suppressor of Fused, Costal-2, Protein kinase A,
Slimb and the CREB-binding protein (Aza-Blanc \& Kornberg 1999, Lefers
{\it et al.}  2001). When SMO is inactive, CI is cleaved to form CIR,
a transcriptional repressor that represses $wg$, $ptc$ (Aza-Blanc \&
Kornberg 1999) and $hh$ transcription (Ohlmeyer \& Kalderon 1998,
M\'ethot \& Basler 1999). When SMO is active, CI is converted to a
transcriptional activator CIA that promotes the transcription of $wg$
and $ptc$ (Alexandre {\it et al.} 1996, von Ohlen \& Hooper 1997,
M\'ethot
\& Basler 1999, Aza-Blanc \& Kornberg 1999).

The gene networks governing embryonic segmentation in {\it Drosophila}
have been modeled using either the discrete-state approach (Bodnar
1997, Bodnar \& Bradley 2001, S\'anchez
\& Thieffry 2001), or the continuous-state approach (Reinitz \& Sharp
1995, Gursky {\it et al.} 2001), but the first modeling work focusing
on the segment polarity gene network was done by von Dassow and
collaborators (von Dassow {\it et al.} 2000, von Dassow \& Odell
2002). von Dassow {\it et al.}  developed a continuous-state model of
the core network of five genes ($en$, $wg$, $ptc$, $ci$ and $hh$) and
their proteins. The initial choice of network topology  failed to reproduce
the observed expression patterns for these genes, but the introduction
of two additional interactions led to surprisingly robust patterning
with respect to variations in the kinetic constants in the rate laws.
The robustness of this model with respect to the changes in the
reaction parameters suggests that the essential features involved are
the topology of the segment polarity network and the signatures of the
interactions in the network (\ie whether they are activating or inhibiting),
and our first objective is to test this prediction.

The Boolean model we develop here is not based on continuous
concentrations of mRNAs and proteins; only two states are admitted for
each component corresponding to whether or not they are present. This
choice is motivated by the ON/OFF character of the
experimentally-observed gene expression patterns\footnote{Indeed,
thresholds are required even in continuous-state models in order to
decide whether a certain concentration corresponds to an ON or OFF
state for comparison with the experimental patterns.}.  We show that
this model reproduces the wild-type gene expression pattern, as well
as the ectopic patterns corresponding to various mutants. We determine
all the steady states of the model analytically, and we find that
there is a surprisingly small number ($6$) of distinct ones, half of
which are observed experimentally. We find the domains of attraction
of these steady states by a systematic search in the space of possible
initial conditions. Furthermore, we determine the minimal
prepatterning necessary to produce the wild-type spatial expression
pattern, and thereby show that the majority of non-initiation errors in the
prepattern can be corrected during the temporal evolution from such
states.  The model demonstrates the remarkable robustness of the
segment polarity network, and shows that the robustness resides in the
topology of the network and the signature of the interactions.

\section{Description of the Boolean model}
\label{rules}

In the model, each mRNA or protein is represented by a node of a
network, and the interactions between them are encoded as directed
edges (see Figure \ref{fig_gene}).\footnote{Note that it is necessary to 
designate different nodes to mRNAs and proteins corresponding to the same gene, 
because in several cases they are expressed in different cells.} The state of each node is $1$ or $0$, 
according as the corresponding substance
is present or not. The states of the nodes can change in time, and the next
state of node $i$ is determined by a Boolean
(logical) function ${\cal F}_i$ of its state and the states of those nodes that
that have edges incident on it. In general, a Boolean or logical function is
written as a statement acting on the inputs using the logical operators "and", 
"or" and "not" and its 
output is $1$($0$) if the statement is true (false). The rules governing the 
transcription of a gene, for example, are determined by a Boolean function of the
state of its transcriptional activators and
inhibitors. Transcription will only commence if the activators are expressed and
the inhibitors are not, thus the Boolean function will have the form "ACT and
not INH", where ACT represents the activators and INH the inhibitors. Moreover,
if this mRNA is translated into a protein, its state enters the Boolean 
function of the protein. This type of network
modeling is rooted in the pioneering work of Kauffman (1969, 1993) on random
Boolean networks and of Thomas 
(1973, Thomas \& D'Ari 1990) on Boolean models describing generic gene
 networks. The novelty of our work
lies in the fact that we base the Boolean model on the known
topology of the {\it Drosophila } segment polarity network.

Expression of the segment polarity gene occurs in stripes that
encircle the embryo, and therefore we can treat the two-dimensional
pattern as one-dimensional. We consider a line of $12$ cells corresponding to three
parasegment primordia (\ie the spatial regions that will become the
parasegments), and impose periodic boundary conditions on the
ends. We use four cells per parasegment primordium because when
expression of the segment polarity genes begins, a given gene is
expressed in every fourth cell. The use of $12$ cells rather than $4$ allows
better illustration of the patterns, but throughout the analysis
we assume that the initial pattern of the segment polarity genes is repeated 
every four cells. We concentrate on the network shown in Figure \ref{fig_gene}, but do not
include FZ and {\it smo} in our base network, because
these nodes are not regulated by other nodes in the network.
Consequently the total number of nodes  in the model system is $n \cdot N = 15\times 12$ 
or $180$.

The states of the nodes evolve in discrete time steps under
the following algorithm. We choose a time interval that is larger or
equal to the duration of all transcription and translation processes,
and we use this interval as the length of a unit timestep. We then
prescribe an initial state for each node, and the state at the next time step 
is determined by the Boolean function ${\cal F}_i$ for
that node.  If $x_{ij}^t $ is the state of the $i^{th}$ node in the $j^{th}$ 
cell at time $t$, and $x = (x_{11}, \cdots, x_{nN})$, then the next state of the
network is $x^{t+1} = {\cal F}(x^t)$, which defines a discrete dynamical system whose iteration
determines the evolution of the state of all nodes. A fixed point of ${\cal F}$ is a time-invariant state
of the system, whereas a fixed point of ${\cal F}^p$ for a minimal $p >1$
represents a state that repeats periodically with least period
$p$. Since the objective is to obtain the
experimentally-observed stable
expression pattern of the segment polarity genes starting from wild-type initial 
conditions, one test of the
model is whether the state evolves to a fixed point that corresponds to this 
pattern.

The Boolean interaction functions are constructed from the
interactions between nodes displayed in Figure \ref{fig_gene}. In order
to focus on the effects of the topology of the segment polarity
network and the signature of the interactions on the predicted steady
states, the model is based on the simplest assumptions concerning the
interactions between nodes. These assumptions can be summarized as
follows\\ 
\indent (i) the effect of transcriptional activators and inhibitors is
never additive, but rather, inhibitors are dominant;

(ii) transcription and translation are ON/OFF functions of the state;

(iii) if transcription/translation is ON, transcripts/proteins are
synthesized in one timestep;

(iv) mRNAs decay in one timestep if not transcribed;

(v) transcription factors and proteins undergoing post-translational
modification decay in one timestep\\
\indent if their mRNA is not present.

For example, EN is translated from $en$, and therefore $EN^{t+1}_i=1$ if
$en^t_i=1$. Since EN is a transcription factor, it is assumed that
its expression will decay sufficiently rapidly that if $en^t_i=0$,
then $EN^{t+1}_i=0$. These two assumptions mean that
$EN^{t+1}_i$ does not depend on $EN^{t}_i$, only on the expression of
$en$, and therefore
\begin{equation}
EN^{t+1}_i=en^t_i.
\end{equation}
\noindent Table \ref{table_rules} gives 
an overview of the Boolean functions for each node, which hereafter are 
labeled by their biochemical symbol. In each case, subscripts signify
spatial position (\ie  cell number) and superscripts signify time.
A detailed rationale for our choice of Boolean rules
used for  updating the 
state of each node is presented in the Appendix.

\begin{table}[htb]
\caption{\small The Boolean functions used in the model. The functions are based on the
known interactions between mRNAs and proteins shown in  Figure
\ref{fig_gene}, and on the temporal assumptions listed above. In general the
updating rule gives the expression of a node at time $t+1$ as a function of the
expression of its effector nodes at time $t$. However, there are three 
exceptions: we assume that the expression of SLP does not change, and that the 
activation of SMO and the binding of PTC to HH are
instantaneous.}
\label{table_rules}
\centerline{
\begin{tabular}{cc}
\toprule
Node & Boolean updating function\\
\midrule
 $SLP_i$ & $SLP_i^{t+1}=SLP_i^t=\left\{\begin{array}{lllll}
0 &\mbox{if}& i\mbox{} \mod4=1 &\mbox{or}& i\mbox{} \mod4=2\\
1 &\mbox{if}& i\mbox{} \mod4=3 &\mbox{or}& i\mbox{} \mod4=0\\
\end{array}\right.$ \\
$wg_i$ & $wg_i^{t+1}=(CIA_i^t$ and $SLP_i^t$ and not $CIR_i^t)$ \\
     & \hspace{3cm} or $[wg_i^t $ and $(CIA_i^t$ or $SLP_i^t )$ and not 
     $CIR_i^t]$ \\
WG$_i$ & $WG_i^{t+1}=wg_i^t$  \\
$en_i$ & $en^{t+1}_i=(WG_{i-1}^t$ or $WG_{i+1}^t)$ and not $SLP^t_i$ \\
$EN_i$ & $EN^{t+1}_i=en^t_i$  \\
$hh_i$ & $hh_i^{t+1}=EN_i^t$ and not $CIR_i^t$ \\
$HH_i$ & $HH^{t+1}_i=hh^t_i$ \\
$ptc_i$ & $ptc_i^{t+1}=CIA^t_i$ and  not $EN_i^t$ and not $CIR^t_i$ \\
$PTC_i$ & $PTC^{t+1}_i=ptc_i^t$ or $(PTC_i^t$ and not $HH_{i-1}^t$ and not 
$HH_{i+1}^t)$ \\
$PH_i$ & $PH^{t}_i=PTC_i^t$ and $(HH^t_{i-1}$ or $HH^t_{i+1})$\\
$SMO_i$ & $SMO_i^t=$ not $PTC_i^t$ or $HH_{i-1}^t$ or $HH_{i+1}^t$ \\
$ci_i$ & $ci_i^{t+1}=$ not $EN_i^t$ \\
$CI_i$ & $CI_i^{t+1}=ci_i^t$ \\
$CIA_i$ & $CIA_i^{t+1}=CI_i^t$ and ($SMO_i^t$ or $hh_{i-1}^t$ or $hh_{i+1}^t)$ \\
$CIR_i$ &  $CIR_i^{t+1}=CI_i^t$ and  not $SMO_i^t$ and not $hh_{i\pm1}^t$\\
\bottomrule 
\end{tabular}}
\end{table}

\section{The functional topology of the segment polarity network}

\begin{figure}[h!]
\centerline{\psfig{figure=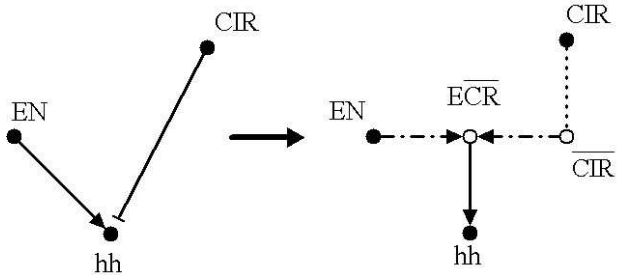,width=.5\textwidth}}
\caption{Illustration of the network expansion process used to construct the
functional topology. To express the logical rule governing the transcription of
{\it hh} graphically, we introduce the complementary node $\overline{\rm{CIR}}$ and
the composite node E$\overline{\rm{CR}}$. The expanded network contains real nodes
(filled circles) and pseudo-nodes (open circles), an inter-dependence relation between
CIR and $\overline{\rm{CIR}}$ (dotted line), edges corresponding to the activation
of E$\overline{\rm{CR}}$ (dash-dotted lines) and a single edge expressing the
activation of {\it hh} transcription.}
\label{fig_expand}
\end{figure}

The rules for advancing the current state of the network given in Table
\ref{table_rules} can be used to construct an expanded graph that reflects the
function of the network. Consider the transcription of the {\it hh} gene. Figure
\ref{fig_gene} shows that {\it hh} has two incoming edges, one from EN and one
from CIR, and Table \ref{table_rules} shows that transcription
of the $hh$ gene requires both the presence of the EN protein and the absence of the
CIR protein. To represent this conjunction in a graphical form, one can say 
that $hh^{t+1}$ depends on the current state of a "pseudo-node"
that we denote by E$\overline{\rm{CR}}$ (see Figure \ref{fig_expand}). The state of 
this new node is one whenever $EN^t=1$ and $CIR^t=0$, and zero otherwise. We can
also represent the dependence of the pseudo-node E$\overline{\rm{CR}}$ on EN and CIR in
a graphical form by introducing a 
"complementary" pseudo-node,
$\overline{\rm{CIR}}$, that is expressed whenever CIR is not. In order to take into
account the inter-dependence of $\overline{\rm{CIR}}$ and CIR we connect them with a
symmetrical (non-directed) edge. Finally, we draw two directed edges starting from EN and
$\overline{\rm{CIR}}$ and ending in E$\overline{\rm{CR}}$, to
represent the dependence of E$\overline{\rm{CR}}$ on the expression of EN and
$\overline{\rm{CIR}}$ (see Figure \ref{fig_expand}).

\begin{table}[htb]
\caption{Definition of the symbols for pseudo-nodes used in Figure
\ref{fig_topology}. The state of each composite node is determined from the 
logical function giving its relation to the state of its "parent" nodes.}
\label{table_notations}
\centerline{
\begin{tabular}{ll}
\toprule
Symbol of pseudo-node & Relation to parent node(s)\\
\midrule
\multicolumn{2}{c} {Complementary nodes} \\
$\overline{\rm{EN}}$ & not EN\\
$\overline{hh}$ & not $hh$\\
$\overline{\rm{HH}}$ & not HH\\
$\overline{\rm{PTC}}$ & not PTC\\
$\overline{\rm{SLP}}$ & not SLP\\
$\overline{\rm{SMO}}$ & not SMO\\
\multicolumn{2}{c} {Composite nodes corresponding to a single cell} \\
(CA$\overline{{\rm ECR}})_2$ & CIA$_2$ and $\overline{{\rm EN}_2}$ and
$\overline{{\rm CIR}_2}$\\
(CAS$\overline{{\rm CR}})_2$ &CIA$_2$ and SLP$_2$ and
 $\overline{\rm{CIR}}_2$\\
(CSM)$_2$ & CI$_2$ and SMO$_2$\\
(E$\overline{\rm{CR}})_2$ & EN$_2$ and $\overline{{\rm CIR}_2}$\\
($w$CA$\overline{{\rm CR}})_2$ & $wg_2$ and CIA$_2$ and $\overline{{\rm CIR}_2}$\\
($w$S$\overline{{\rm CR}})_2$ & $wg_2$ and SLP$_2$ and $\overline{{\rm CIR}_2}$\\
\multicolumn{2}{c} {Composite nodes corresponding to intercellular 
interactions} \\
C$_ih_j$ & CI$_i$ and $hh_j$\\
C$_i\overline{{\rm SM}_ih_jh_k}$ & CI$_i$ and
$\overline{{\rm SMO}}_i$ and $\overline{hh_j}$ and
$\overline{hh_k}$\\
P$_i\overline{{\rm H}_j\rm{H}_k}$ & PTC$_i$ and $\overline{{\rm HH}_j}$ and
 $\overline{{\rm HH}_k}$\\
P$_i$H$_j$ & PTC$_i$ and HH$_j$\\
W$_i$S$_j$ & WG$_i$ and SLP$_j$\\
\bottomrule
\end{tabular}}
\end{table}

\begin{figure}[htb]
\centerline{\psfig{figure=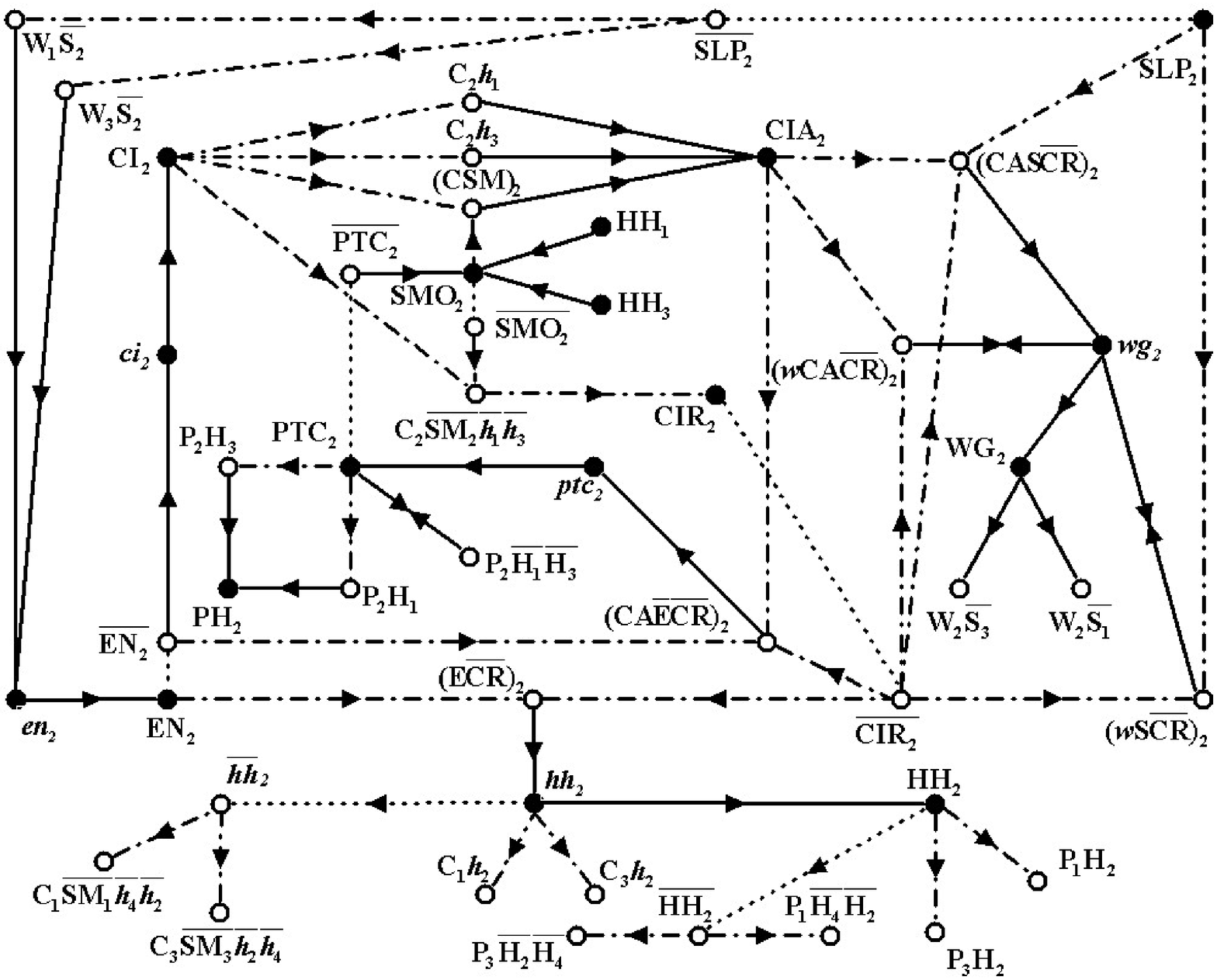,width=.65\textwidth,angle=-90}}
\caption{Functional topology of the network affecting the second cell of 
the parasegment. Filled nodes represent real nodes, open circles represent 
pseudo-nodes, denoted according to Table
\ref{table_notations}. Pseudo-nodes with multiple indexes correspond
 to intercellular interactions and either receive some of their inputs
 from the neighboring cells, or contribute to the expression of the
 nodes in the neighboring cells (not shown). Symmetrical edges between nodes
 and their complementaries are drawn with dotted lines; edges determining the 
 expression of composite nodes are drawn with dash-dotted lines; edges
 determining the expression of real nodes in the next timestep are continuous.
 Double arrows denote a pair of oppositely directed edges.}
\label{fig_topology}
\end{figure}

In general we define "complementary" pseudo-nodes for any
nodes whose negated expression enters Table \ref{table_rules}, and
"composite" pseudo-nodes for any combinations of node expressions that are connected
by the "and" operator in Table \ref{table_notations}. In effect, we expand the initial network
into a kind of bipartite network composed of real and pseudo-nodes; the real nodes are the
original nodes while the pseudo-nodes correspond to the regulation rules between the
nodes\footnote{In regular bipartite networks only edges connecting different kinds
of nodes are allowed, however, our network contains edges between nodes of the
same kind.}. The total number of nodes in the network for $12$
cells is $37\times 12=444$. Of the $37$ nodes per cell, $28$ are cell-autonomous, and
$9$ composite nodes represent signaling between neighboring cells. Figure
\ref{fig_topology}  shows the nodes and edges corresponding to the mRNAs and proteins
in the second cell of the parasegment ($15$ real nodes, drawn as filled circles) 
together with the pseudo-nodes these mRNAs and proteins influence, both cell-autonomously ($7$
complementary and $6$ composite pseudo-nodes, drawn as open circles), and in the
neighboring cells ($12$ composite pseudo-nodes, drawn as open  circles). To
illustrate the  symmetrical nature of the intercellular signaling, we also include on
the figure the  pseudo-nodes in neighboring cells that affect the expression of mRNAs
and proteins in the second cell ($6$ composite pseudo-nodes, drawn as open
circles).

Although the introduction of the pseudo-nodes increases the number of nodes in
the network, it eliminates the distinction between edges based on their
signatures; all directed edges in Figure \ref{fig_topology} now signify
activation. However, there are differences in the way  multiple
activating edges are taken into account: multiple edges ending in composite
pseudo-nodes (dash-dotted lines in Figure \ref{fig_topology}) are added 
by the operator "and", while multiple edges starting at pseudo-nodes and ending in
real nodes are cumulated by the operator "or". The presence of inhibitory
control in Figure \ref{fig_gene} is reflected in activation by complementary 
nodes in Figure \ref{fig_topology}.
Although all directed edges signify activation, there
are temporal differences in the underlying processes. The expression of the
pseudo-nodes at time $t$ is set by the expression of their parent
nodes at time $t$, while the expression of most real nodes at time $t$ is set
by the expression of pseudo- and/or real nodes at time $t-1$.

 Figure \ref{fig_topology} illustrates the
heterogeneous functional topology of the segment polarity network. The
majority of nodes have few edges, but there are key nodes with a
large number of incoming or outgoing edges. For example,
$\overline{\rm{CIR}}$ has $5$ outgoing edges, while HH has $4$. The
important role of HH in the network is reflected in the fact that it
affects the future expression of $4$ other proteins (CIA, CIR, PTC and
PH) in the neighboring cells, \ie $8$ nodes.  Other nodes such as
CIA have several incoming edges, indicating that they can be activated
in many ways. A single node,
SLP, has only outgoing edges because it is constitutively present; all
others have both incoming and outgoing edges (the apparent
exceptions interact with nodes in the neighboring cells).

The functional network of Figure \ref{fig_topology} gives  insight into the
time-evolution of the expression of the segment polarity genes. Given an initial
state for all $15$ real nodes in the network at a time $t$, the state of the 
complementary nodes is the negation of
the state of the corresponding real node. The state of a composite pseudo-node 
is $1$ if all of its parent nodes have state 
$1$, otherwise it is $0$.
The state of real and pseudo-nodes at time $t$ determines the state of real nodes at time 
$t+1$. If a real node has multiple incoming edges, its state
is $1$ whenever at least one node at the origin of these edges is expressed.
This way the general dynamical system $x^{t+1} = {\cal F}(x^t)$ can be decomposed into
\begin{equation} 
x^{t+1}=T_{or}(T_{and}(x^t)), 
\end{equation} 
where the operators $T_{or}$ and $T_{and}$ correspond to the updating processes
described above. It is tempting to imagine these operators as matrices, but
unfortunately the "and" and "or" operators cannot be linearized (i.e., "A and B"
cannot be written as a linear combination of A and B).

While it is not possible to follow the evolution of the network's
expression analytically, we can gain insight into it by studying the
cluster (group) of nodes that can be activated by the expression
of a given node. We can estimate the size of this cluster by
successively following the directed edges starting from the node, then
those starting from the endpoints of these edges and continuing until
the edges leave Figure \ref{fig_topology}.  This method gives us an
upper bound for the number of activated nodes, since in most cases
additional conditions must be satisfied in order for a node to be
activated. Note that in this method we cannot follow the symmetrical
edges, since their endpoints have opposite expressions. The absence of
EN (or conversely the presence of $\overline{\rm{EN}}$) gives the
largest activated cluster, containing {\it ci}, CI, CIA, {\it
ptc}, PTC, PH, {\it wg}, and WG. A separate activated cluster starts
with the presence of {\it en}, and contains EN, {\it hh } and
HH. These activating clusters indicate that the stripe of {\it en} and
{\it hh} never overlaps with those of $wg$, $ptc$ and $ci$. This
separation into anterior and posterior compartments expressing
different genes is well-known, in fact, it is the basis
for calling these genes "segment polarity genes" (Wolpert {\it et al.} 1998).

While the majority of the activating effects propagate outside the
cell, there are three cases in which an activation can return to its
source. In other words, three short cycles exist in the network of
Figure \ref{fig_topology}. The first two cycles connect $wg_2$ with
$(w{\rm CA}\overline{{\rm CR}})_2$ or $(w{\rm S}\overline{{\rm
CR}})_2$ and the third connects PTC$_2$ with P$_2\overline {{\rm
H}_1{\rm H}_3}$. These cycles ensure the maintenance of {\it wg} and
PTC if all the conditions for the expression of the pseudo-node in the
cycle are met (i.e. {\it wg} is maintained if SLP or CIA is present
and CIR is absent, PTC is maintained if HH is not expressed in the
neighboring cells). The successful activation of the {\it wg} cycle
can induce the stable expression of {\it en} and {\it hh} in those
neighboring cells where neither SLP nor CIR is expressed, and stable
expression of PTC leads to stable CIR expression two cells removed from
{\it en} expression. The results presented in the following sections
confirm the special role of these cycles in the dynamics of the
segment polarity gene expression pattern.

\section{Comparison between numerical and experimental results}
\label{sect_valid}

 The segment polarity genes are activated by the pair-rule genes in
the cellular blastoderm phase (stage $5$ according to the classification
of Campos-Ortega \& Hartenstein 1985) of the {\it Drosophila}
embryogenesis, and maintain the parasegment borders and later the
polarity of the segments from the end of gastrulation through
germ-band elongation (stages $8$-$11$, see Wolpert {\it et al.} 1998). 
The parasegment borders 
form between the $wg$ and $en/hh$ expressing cells 
(Hooper \& Scott 1992, Wolpert {\it et al.} 1998).  
Since our model is intended to describe the effect of the segment polarity genes in
maintaining the parasegment border, the patterns of segment polarity
genes formed before stage $8$ can be considered as given initial states, and 
the final stable state should coincide with the wild-type
patterns maintained during stages $9$-$11$.

\begin{figure}[htb]
\centerline{\psfig{figure=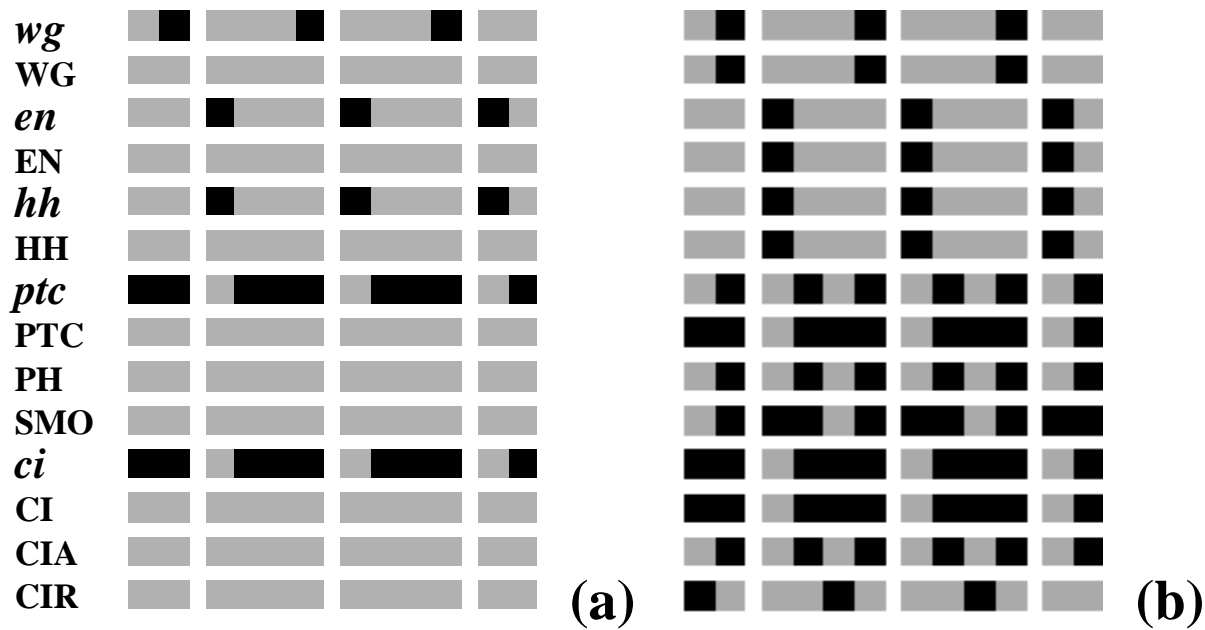,width=.5\textwidth,angle=-90}}
\caption{Wild-type expression patterns of the segment polarity genes. Here and
hereafter left corresponds to anterior and right to posterior in each
parasegment. Horizontal rows correspond to the pattern of individual nodes
- specified at the left side of the row - over two full and two partial 
parasegments. Each parasegment 
is assumed to be four cells wide. A black (gray) box denotes a node that is
ON (OFF). (a) The experimentally-observed 
initial state before stage $8$. {\it en, wg}
and {\it hh} are expressed in one-cell-wide stripes, while the broad {\it ptc} 
and {\it ci} stripes are complementary to {\it en}. (b) The steady state of the 
model when initialized with the pattern in (a). This pattern is in agreement with
the observed gene expression patterns during stages $9$-$11$, see text.}
\label{fig_wild}
\end{figure}

The initial state of each parasegment primordium, based on the experimental 
observations of stage $8$
embryos, includes a two-cell-wide SLP stripe in the posterior half 
(Cadigan {\it et al.} 1994), a single-cell-wide $wg$ stripe in
the most posterior part (Hooper \& Scott 1992), 
single-cell-wide
$en$ and $hh$ stripes in the most anterior part (Tabata {\it et al.} 1992, 
Hooper \& Scott 1992), and $ci$ and $ptc$ expressed in the posterior
three-fourths (Hidalgo \& Ingham 1990, Hooper \& Scott 1992, 
Taylor {\it et al.} 1993). Since
the proteins are translated after the mRNAs are transcribed, we assume
that the proteins are not expressed in the initial state. The one-dimensional
representation of the mRNA and protein patterns is shown in
Figure \ref{fig_wild}(a).

We iterate the dynamical system defined by the rules in Table \ref{table_rules}
starting from the initial state described above. We find that after only $6$ time
steps, the expression pattern stabilizes in a time-invariant spatial pattern
(see Figure \ref{fig_wild}(b)) that coincides with the experimentally
observed wild-type expression of the segment polarity genes
during stages $9$-$11$. Indeed, $wg$ and WG are expressed in the most posterior cell
of each parasegment (Ingham {\it et al.} 1991), while  $en$, EN, $hh$ and HH are expressed
in the most anterior cell of each parasegment, as is observed experimentally
(Ingham {\it et al.} 1991, Tabata {\it et al.} 1992), $ptc$ is expressed in two stripes of cells, one stripe on each side
of the $en$-expressing cells, the anterior one coinciding with the $wg$ stripe
(Hidalgo \& Ingham 1990, Hooper \& Scott 1992). SMO is present in broad stripes
whose anterior border coincides with the anterior border of the $wg$ stripe and
whose posterior border extends about one cell further from the $en$ stripe
(Alcedo {\it et al.} 2000). $ci$ is expressed almost ubiquitously, with the
exception of the cells expressing $en$ (Eaton \& Kornberg 1990, Alexandre {\it
et al.} 1996). CIA is  expressed in the neighbors of the HH-expressing cells,
while CIR is expressed far from the HH-expressing cells (Aza-Blanc \& Kornberg 
1999). Note
that while the majority of the proteins are expressed in the same cells as
their mRNAs, this is not the case for PTC. Indeed, while the $ptc$ stripe
separates into two by stage $11$ (Taylor {\it et al.} 1993), the PTC stripe 
remains broad.
There are indications that the level of PTC decreases in the middle of the
stripe (Taylor {\it et al.} 1993), but the existence of PTC in those cells is very important
because their signaling maintains the production of CIR (see Figure
\ref{fig_wild}(b)).

We have also done a systematic analysis of the patterns obtained when the
initial expression of individual genes or groups of genes differs from
the wild-type initial condition. In principle the attractor for some
initial conditions could be periodic in time, but we have found that
the only stable attractors are steady states. Since the purpose of the
segment polarity network is to stabilize and maintain the parasegment
borders, this result is biologically realistic.

\begin{figure}[h!]
\centerline{\psfig{figure=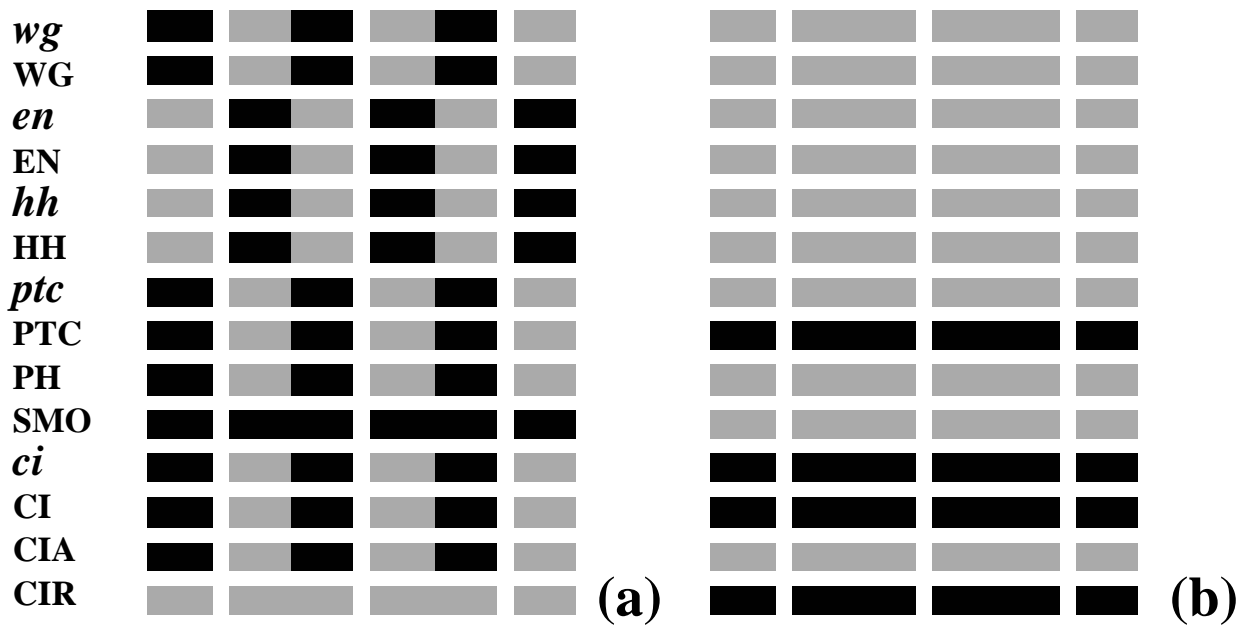,width=.5\textwidth,angle=-90}}
\caption{Ectopic expression patterns of the segment polarity genes
obtained from the model by varying the initial conditions or inactivating
certain nodes. (a) Broad type expression pattern. The
stripes of $en$, $wg$, $ptc$ and $hh$ are broader than normal, while the $ci$
stripe narrows and CIR is not expressed. The anterior broadening of the $en$
stripe together with the posterior broadening of the $wg$  stripe induces an
ectopic "border" in the middle of the parasegment. This state arises if $wg$, 
$en$ or $hh$
 is initiated in broader stripes than wild type. This pattern is in
 perfect agreement with the experimentally-observed gene expression after heat
 shock experiments on $en$ and $hh$ (see text). A similar pattern, only
 without $ptc$, PTC and PH expression, is obtained from the model when the expression of $ptc$
 is kept OFF, in agreement with observations on $ptc$ mutants (see text).
(b) Stable pattern with no stripes for $wg$, $en$, $hh$ and $ptc$. This pattern
 arises if any of $wg$, $en$ or $hh$ is kept OFF in the model, when $wg$
 initiation is substantially delayed, or when 
 intercellular interactions are disrupted.}
\label{fig_mutant}
\end{figure}

We first concentrate on the overexpressed initial patterns, as these
provide direct comparison with heat-shock experiments. In these experiments 
the initial expression of a selected gene is
ubiquitously induced following a heat shock. According to these
experiments, the $wg$ and $ptc$ stripes expand anteriorly when $hh$ is
ubiquitously induced (Gallet {\it et al.} 2000). When the same induction is 
done on
$en$, broadened $en$ stripes result (Heemskerk {\it et al.} 1991), and
narrower $ci$ stripes emerge after a transient decay of $ci$ (Schwartz {\it et al.}
1995). Our model indicates that these two cases lead to the
same steady state expression pattern that incorporates all experimental
observations: broad $en$, $wg$, $ptc$ and $hh$ stripes, and narrower
$ci$ stripes (see Figure \ref{fig_mutant}(a)). We can understand the
process leading to this state from the functional topology of the
network (Figure \ref{fig_topology}): ubiquitous {\it hh} means that
$\overline{hh}$ is not expressed in any of the cells, nor is any
C$_i\overline{{\rm SM}_i h_j h_k}$, thereby removing the only path to
the production of CIR. This means that CIA is produced in each {\it
ci} -expressing cell, and it activates the transcription of {\it wg} and
{\it ptc} in a wider domain.

Another method for inferring gene interactions experimentally is to
silence selected genes by mutations. These inactive genes can be simulated by setting
the expression of the transcript to zero and not updating it during
the evolution of the system. Our results indicate that if any of
$en$, $wg$ or $hh$ are blocked, while the other genes are
initiated in the wild type pattern (see Figure \ref{fig_wild}(a)), the steady
state is a pattern with no $en$, $wg$, $ptc$ or $hh$, as in
Figure \ref{fig_mutant}(b). We can see from Figure \ref{fig_topology} that
each of these mutations disrupts intercellular signaling, causing
ubiquitous expression of CIR, which in turn leads to ubiquitous
repression of transcription. In agreement with this result, it has
been observed experimentally that the $hh$ expression in $en$ null
embryos starts normally, but disappears before stage $10$
(Tabata {\it et al.} 1992). In $wg$ null embryos, $en$ is initiated normally but
fades away by stage $9$, as observed by DiNardo {\it et al.} (1988),
 while $ci$ is ubiquitously expressed (Schwartz {\it et al.} 
1995). In $hh$ mutant embryos the $wg$ expression disappears by stage $10$ (Hidalgo \& Ingham 
1990), as does the expression of $ptc$, and there is no segmentation 
(Gallet {\it et al.} 2000). All these experimental results are in excellent agreement with the 
 numerically obtained pattern shown in Figure \ref{fig_mutant}(b).

If the $ptc$ gene is blocked, we obtain a pattern with broad $wg$, $en$ and $hh$
stripes, which differs from the pattern of Figure \ref{fig_mutant}(a) only in the fact that
there is no $ptc$, PTC and PH expression. Indeed, the network in Figure \ref{fig_topology} shows 
that the role of $ptc$ is to restrict ectopic expressions of $en$, $wg$ and 
$hh$; if $ptc$ is deactivated, $\overline{PTC}$ will be ubiquitous, causing all
CI to be transformed into CIA, which leads to  broadening of the
$wg$ and $en/hh$ stripes. This pattern agrees with the experimental results on
 $ptc$ mutants. These results indicate broad $en$, $wg$ and $hh$ stripes 
 (Tabata {\it et al.} 1992, Gallet {\it et al.}
 2000); Martinez-Arias {\it et al.} (1988) and  Gallet {\it et al.} (2000) 
 find that a new ectopic groove forms at the second $en-wg$ interface at the middle of the
parasegment. Also, $ci$ is not expressed at this ectopic groove
(Schwartz {\it et al.} 1995). Moreover our results are in agreement with all
experimental observations of double mutants as well (DiNardo {\it et al.} 1988,
Ingham  {\it et al.} 1991, Tabata {\it et al.} 1992, Bejsovec and Wieschaus 1993).

If all the other genes are initiated normally, we find that the effect
of a $ci$ deletion does not affect the $en$, $wg$ and $hh$
patterns. However, $ptc$ expression disappears, while PTC is expressed
in a single cell wide stripe. Indeed, Figure \ref{fig_topology} shows
that the deactivation of $ci$ leads to the disappearance of CIA and
CIR, but wild-type $wg$ can still be maintained by SLP, and the
interactions between $wg$, $en$ and $hh$ can maintain the $en/hh$
stripe as well. In this case experiments indicate that the segmental
grooves are present and $wg$ is expressed until stage $11$, but $ptc$
expression decays (Gallet {\it et al.} 2000). Our results confirm the 
experimental result that the deletion of $ci$ is able to counter the broadening
effect of heat-shock-induced ubiquitous $hh$ expression (von Ohlen \& Hooper 
1997). Correct segmentation is maintained even for the other extreme, in
$ci$ - $hh$ double mutants, in agreement with the results of Gallet {\it et al.}
(2000). These findings indicate that the role of $ci$ is to provide additional
control over a functional sub-network formed by $en$, $wg$, $hh$ and SLP
This sub-network can maintain an initial wild-type pattern, but loses
expression in the case of initiation errors. The existence of the dual
transcription factors CIA and CIR suggests a mechanism to correct both under-
and overexpression errors in the initiation of $wg$ and $en$.

\section{Determination of the steady states and their domains of attraction}

The fact that the Boolean model reproduces the
results of numerous experiments remarkably well suggests that
the structure of the model is essentially correct, and
warrants exploration of problems that have not been studied
experimentally. For example, we can determine the complete set of
stable steady state patterns of segment polarity gene expression, and
estimate the domain of attraction of these states. The former can be
done analytically by noting that
these are fixed points of the discrete dynamical system, and so
$x^{t+1}_i=x^t_i$. Thus a steady state is the solution of the system of equations
obtained from Table \ref{table_rules} by simply removing  the time
indices. This leads to the following equations. 
\begin{eqnarray}
\label{eq_stable}
SLP_i&=&\left\{\begin{array}{lllll}
0 &\mbox{if}& i\mbox{} \mod4=1 &\mbox{or}& i\mbox{} \mod4=2\\
1 &\mbox{if}& i\mbox{} \mod4=3 &\mbox{or}& i\mbox{} \mod4=0\\
\end{array}\right.\nonumber\\ 
wg_i&=&(CIA_i \mbox{ and } SLP_i \mbox{ and not } CIR_i) \mbox{ or }  
[wg_i \mbox{ and } (CIA_i \mbox{ or } SLP_i ) \mbox{ and not }
CIR_i]\nonumber\\ 
WG_i&=&wg_i\nonumber\\  
en_i&=&(WG_{i-1} \mbox{ or } WG_{i+1}) \mbox{ and not } SLP_i\nonumber\\ 
EN_i&=&en_i\nonumber\\ 
hh_i&=&EN_i \mbox { and not } CIR_i\nonumber\\
HH_i&=&hh_i\nonumber\\ 
ptc_i&=&CIA_i \mbox{ and  not } EN_i \mbox{ and not } CIR_i \\ 
PTC_i&=&ptc_i \mbox { or } (PTC_i \mbox { and not } HH_{i-1} \mbox{ and not } 
HH_{i+1})\nonumber\\ 
PH_i&=&PTC_i \mbox { and } (HH_{i-1} \mbox { or } HH_{i+1})\nonumber\\ 
SMO_i&=&\mbox { not } PTC_i \mbox { or } HH_{i-1} \mbox { or }
HH_{i+1}\nonumber\\
ci_i&=&\mbox{not } EN_i\nonumber\\ 
CI_i&=&ci_i\nonumber\\ 
CIA_i&=&CI_i \mbox { and } ( SMO_i \mbox { or } hh_{i-1} \mbox { or } hh_{i+1})\nonumber\\ 
CIR_i&=&CI_i \mbox { and not } SMO_i \mbox { and not } hh_{i-1} \mbox { and not }
hh_{i+1}\nonumber 
\end{eqnarray}

\begin{figure}[htb]
\centerline{\psfig{figure=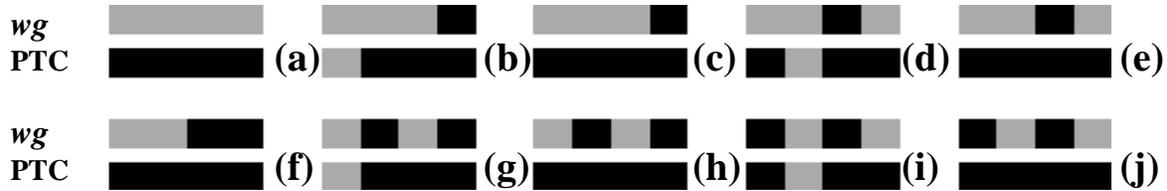,width=.15\textwidth,angle=-90}}
\caption{Patterns of {\it wg} and PTC obtained as solutions of Equation 
(\ref{eq_sol}). (a) Solution leading to the non-segmented pattern of
 Figure \ref{fig_mutant}(b). (b) Solution corresponding to the wild-type pattern
 of Figure \ref{fig_wild}(b)  (c) Solution leading to a variant of the wild-type
 pattern of Figure\ref{fig_wild}(b) differing from it only in the expression of 
 PTC that in this case becomes ubiquitous. (d), (e) Ectopic solutions that lead to patterns with no
parasegment borders. The only difference between the two patterns is in the
width of the PTC stripe ($3$ and $4$-cells-wide, respectively). (f) Solution
leading to the broad type pattern of Figure
\ref{fig_mutant}(a). (g), (h) Almost wild-type solutions with two $wg$ stripes.
The two patterns differ only in the width of the PTC stripe.
(i), (j) Ectopic solutions similar to (d) and (e), but with two $wg$ stripes. }
\label{fig_sol}
\end{figure}

Since the parasegments are assumed to be identical, 
we can restrict attention to one parasegment, and assume that $i$ can run only 
from $1$ to $4$, i.e. the width of one parasegment. The majority of the 
variables in Equation \ref{eq_stable} can be 
eliminated, with the exception of $wg_i$ and PTC$_i$. The expression of these
nodes appears on both sides of the equations, reflecting the cycles of 
{\it wg} and PTC in Figure \ref{fig_topology}. The final set of
equations to be solved to find the steady states is
\begin{eqnarray}
\label{eq_sol}
wg_1&=& wg_1 \mbox { and not } wg_2 \mbox{ and not } wg_4\nonumber\\
wg_2&=& wg_2 \mbox { and not } wg_1 \mbox{ and not } wg_3\nonumber\\
wg_3&=& wg_1 \mbox { or } wg_3\nonumber\\
wg_4&=& wg_2 \mbox { or } wg_4 \\
PTC_1&=& (\mbox { not } wg_2 \mbox { and not } wg_4) \mbox{ or } (PTC_1 
\mbox { and not } wg_1 \mbox{ and not } wg_3)\nonumber\\
PTC_2&=& (\mbox { not } wg_1 \mbox { and not } wg_3) \mbox{ or } (PTC_2 
\mbox { and not } wg_2 \mbox{ and not } wg_4 )\nonumber\\
PTC_3&=&1\nonumber\\
PTC_4&=&1\nonumber
\end{eqnarray}
The asymmetry of these equations with respect to the anterior and
posterior parts of the parasegments is due to the effect of SLP.
Note that this set of equations shows that specifying compatible expressions for 
{\it wg} and PTC is sufficient for the complete description of a stable
pattern. We solve this system of equations by examining all $2^6=64$ possible
states to determine whether they remain unchanged under the
transformations given in  Table \ref{table_rules}. We
obtain $10$ solutions, pictured on Figure \ref{fig_sol}. The complete patterns can be obtained from the solutions on Figure \ref{fig_sol} by
using Equation (\ref{eq_stable}) to determine the expression of the other nodes 
in the network.

The first steady state, resulting from solution
\ref{fig_sol}(a), is the pattern with no segmentation first presented in Figure
\ref{fig_mutant}(b). The solution \ref{fig_sol}(b) corresponds to the wild-type 
pattern shown in Figure \ref{fig_wild}(b). There is also a variant of this solution,
shown on Figure \ref{fig_sol}(c), that has ubiquitous PTC expression rather 
than
the posterior three quarters of the parasegment. This solution leads to a pattern
that differs from the wild-type pattern only in the expression of PTC, thus we do
not classify it as a distinct steady state. 

\begin{figure}[htb!]
\centerline{\psfig{figure=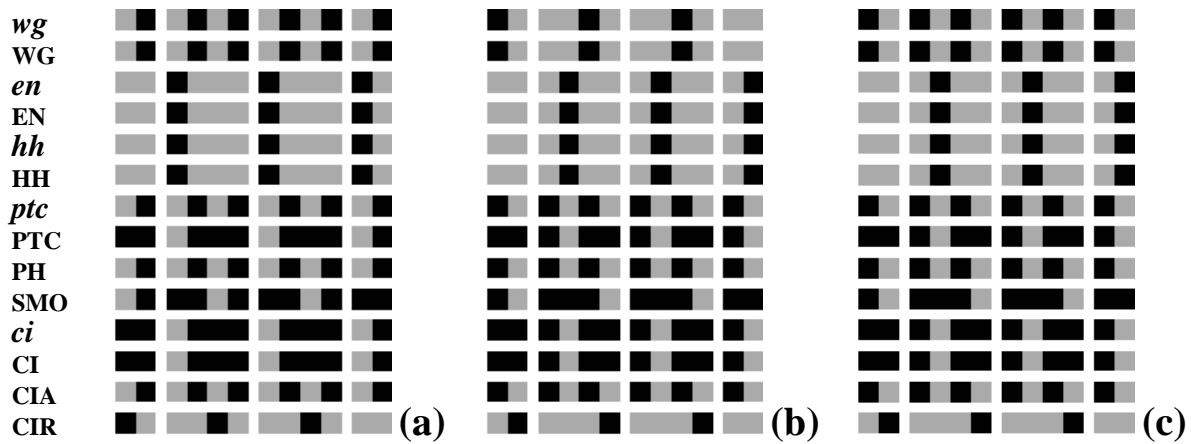,width=.35\textwidth,angle=-90}}
\caption{Various stable patterns of the segment polarity genes obtained from
 Equation (\ref{eq_stable}). (a) Almost wild-type type expression pattern
with two $wg$ stripes. This state corresponds to the solution of Equation
\ref{eq_stable} presented on Figure
\ref{fig_sol}(g). (b) Ectopic pattern with
displaced $wg$, $en$ and $hh$ stripes. This state arises from the solution on
Figure \ref{fig_sol}(d). (c) Variant of the ectopic
pattern shown in (b) with two $wg$ stripes. This state is determined from the
solution \ref{fig_sol}(h).}
\label{fig_stable}
\end{figure}

In the third distinct steady state,
corresponding to the solution \ref{fig_sol}(d), the $wg$ stripe is displaced
anteriorly from its wild-type position, while the $en/hh$ stripe is displaced
posteriorly (see Figure\ref{fig_stable}(b)). This expression pattern corresponds to
an ectoderm with no parasegmental grooves, since the end of the $wg$ stripe does not
meet the beginning of the $en$ stripe. The solution \ref{fig_sol}(e) corresponds to
the ubiquitous PTC variant of this pattern. The solution with broadened $wg$ and PTC stripes
shown in Figure \ref{fig_sol}(f)
leads to the broad type pattern first presented on Figure \ref{fig_mutant}(a).

The next solution, \ref{fig_sol}(g), leads to a steady state that differs  from the wild
type only in the fact that it contains two $wg$ stripes, one in its normal position
and one in a symmetrical position in the anterior part of the parasegment. The
solution \ref{fig_sol}(h) leads to a variant with ubiquitous PTC. The last
distinct steady state, corresponding to the solution \ref{fig_sol}(i), is similar to the
ectopic pattern of Figure \ref{fig_stable}(b), but it has two $wg$ stripes, and an
ectopic parasegment border between the third and fourth cell of the parasegment (see
Figure\ref{fig_stable}(c)). The solution \ref{fig_sol}(j) leads to a ubiquitous PTC
variant of this state. 

In summary, the $10$ solutions of Equation \ref{eq_sol} lead to six distinct 
steady states. Three of these steady states are well-known experimentally,
corresponding to the wild-type pattern and two mutant patterns with either no
stripes or broadened stripes. The existence of three additional states not
observed experimentally suggests that the network can also produce
patterns beyond those required in  {\it Drosophila} embryogenesis. One of
these states, corresponding to solution \ref{fig_sol}(g), coincides with the wild-type steady state in the expression of all
nodes except for $wg$, but this divergence from wild type implies the emergence
of a second, ectopic parasegment border. The second ectopic steady state
(Figure \ref{fig_stable}(b)) has
displaced stripes for all nodes, and no parasegment borders, while the
last ectopic state is a two-$wg$  variant of this state with ectopic parasegment
borders. None of these patterns have been observed experimentally,
probably because they require ectopic initial conditions (see below).
Additionally, if  we relax the assumption of identical parasegments, the steady
state patterns corresponding to individual parasegments can  be combined to
form diverse patternings of the whole ectoderm. For example, the wild-type
steady state and its double $wg$ variant can be seamlessly integrated.

While each of the steady states can be obtained starting from suitable
nearby states, the number of initial conditions leading to a chosen
stable pattern, {\it i.e.}, its domain of attraction, can be very
different.\footnote{A systematic method of computing state pre-images in random
Boolean networks has been proposed by Andrew Wuenche (1997).} 
Consider first the number of
initial states that lead to the wild-type steady state. If we fix all
nodes but one in their wild-type pattern, there are $2^4=16$ distinct
initial patterns corresponding to the four cells of the
parasegment. We do this for each of the $14$ variable nodes in turn (we do not
change the expression of SLP) and find that the number
of initial patterns leading to the wild-type steady state is $3$ for
$wg$ or WG variation, $4$ for $en$, EN, $hh$ or HH variation, $8$ for
$ptc$, PTC, CI or CIA variation, and $16$ for PH, SMO, $ci$ or CIR
variation. We then use initial states in which the expression of pairs and 
triplets of nodes is different from the wild-type pattern. We find that the
wild-type steady state is reachable from all
initial conditions for which {\it each variable node} has one of the patterns 
that were previously found to lead to wild type when {\it only the expression 
of that node was varied}. If we now extend this result and assume that the attractor is the wild-type
pattern if the prepattern for each node taken separately is one
that led to wild-type expression when only the pattern of that node 
was varied, the number of network-wide
wild-type inducing prepatterns can be calculated by multiplying the numbers of
individual wild-type inducing prepatterns \ie\ $n_{wg}\times n_{WG}...\times
n_{CIR}$, where $n_{x}$ is the number of wild-type inducing prepatterns of
node $x$.  There are $3^2\cdot4^4\cdot8^4\cdot16^4\sim 6\times 10^{11}$ 
such prepatterns for each parasegment, which is a fraction of $8\times 10^{-6}$ of the
total number of initial states $N_{st}=16^{14}$. 

This fraction seems small, but we should not forget that not all
initial conditions are biologically realistic. Indeed, it is almost impossible
to initiate $wg$, $en$ or $hh$ in a broad stripe as their initiation depends on
complex interactions between different pair-rule gene products (see Wolpert {\it et al.
} 1998). It is much more realistic to focus on under-expression errors
corresponding, for example, to delays in initiation. We find that the network is 
very robust with respect to missing initial expression of nodes. We have 
determined that the minimal prepatterning that leads to wild-type stable 
expression is as follows.

{\bf 
\noindent $\bullet$ $wg$ is wild type,\\
$\bullet$ $en$ and $hh$ are not expressed,\\
$\bullet$ $ptc$ is expressed in the third cell of the parasegment primordium,\\
$\bullet$ $ci$ and the proteins are not expressed.\\ }
\noindent In summary, it is enough to initiate the expression of two genes in
two cells per parasegment primordium, and the interactions between the segment polarity
genes will initiate the others. While this result might not be completely
realistic due to the assumptions of our model, it suggests a remarkable 
error-correcting ability for the segment polarity gene network.

Note that the minimal prepattern contains the wild-type stripe of $wg$. If 
$wg$ is not expressed initially we find that the final pattern has no stripes,
as shown in Figure \ref{fig_mutant}(b), regardless of the initial
pattern of the other nodes. Consequently, a fraction of at least
$1/16^{th}$ of the initial states leads to the pattern of
Figure \ref{fig_mutant}(b). This finding suggests that $wg$ has a
special role in the functioning of the segment polarity network, and
has to be activated at a specific time and specific cells in order
to obtain wild-type gene expression.
 
In the other limit, broader than wild-type initial expression of any node 
except PH, SMO, $ci$ and CIR leads to the pattern with broad stripes as in
Figure \ref{fig_mutant}(a). This pattern is obtained in the vast majority of
prepatterns, comprising about $90\%$ of the total number of initial states.

The almost wild-type steady state pattern with two $wg$ stripes can
only appear if CIA is initialized at the same time as $wg$, and $wg$
is prepatterned with this ectopic pattern. Since the activation of CI
to CIA requires HH signaling, and the HH stripes were found to appear
later than the $hh$ stripes (Taylor {\it et al.} 1993), the probability of
simultaneous $wg$ and CIA pre-expression is small. 

The minimal prepattern needed for the ectopic pattern with
displaced $wg$ and $en$ stripes (Figure \ref{fig_mutant}(c)) is $wg$
expression in the third cell of the parasegment primordium (the same as its
steady pattern), and $ptc$ expression in the last cell of the
parasegment primordium, where the wild-type stripe of $wg$ would normally
be. Note that this minimal initial condition is simply a shifted
version of the minimal condition for the wild-type steady state.
Indeed, the number of initial patterns leading to this steady state is
the same as the number of wild-type inducing initial states, {\it
i.e.}  a fraction of $8\times 10^{-6}$ of the total number of initial
states. In practice the simultaneous ectopic initiation of several
nodes is very improbable, and indeed, this steady state has never been
observed. The variant with two $wg$ stripes is even more improbable,
requiring simultaneous ectopic prepatterning for $wg$, $ptc$ and CIA. Finally,
all ubiquitous-PTC variants of distinct steady states (corresponding to the 
solutions \ref{fig_sol}(c),\ref{fig_sol}(e), \ref{fig_sol}(h) and
\ref{fig_sol}(j)) require a ubiquitous PTC prepattern in addition to the
prepattern required for their striped-PTC counterparts.

\section{A two-step model}

While the steady states of the model reproduce experimentally-observed
expression patterns, the temporal evolution may not reflect the {\em
in vivo} evolution of expression patterning. We assume that the expression of 
mRNAs/proteins decays in one time step if
their transcriptional activators/mRNAs are switched off, and this assumption
can induce transient on-off flickering in the expression
pattern. Such flickering can be eliminated by relaxing the
assumptions of immediate switch-off, and assume more realistic, longer decay times.
 
As an illustration we consider a variant of our model incorporating slower decay of
proteins. In this model we assume that protein expression in the $(t+1){st}$ step
depends on the expression of their mRNAs in the previous two steps,
$t$ and $t-1$. According to this assumption, if the transcript is not 
present at time $t$, but it was
expressed at time $t-1$, we assume that the protein's expression
(initiated at time $t$) is maintained at time $t+1$. In other words,
the expression of a protein is maintained for at least two steps
following the appearance of the transcript. The minimum value of two
steps is obtained when the transcript appears and then
disappears. According to the two-step assumption, the logical rules
describing the expression of EN, WG, CI and HH become
\begin{eqnarray}
EN_i^{t+1}&=&en_i^t \mbox{ or } en_i^{t-1}\nonumber\\ 
WG_i^{t+1}&=&wg_i^t \mbox{ or } wg_i^{t-1}\nonumber\\
CI_i^{t+1}&=&ci_i^t \mbox{ or } ci_i^{t-1}\nonumber\\
HH_i^{t+1}&=&hh_i^t \mbox{ or } hh_i^{t-1}\nonumber
\end{eqnarray}
The logical rule governing the expression of PTC (Equation \ref{eq_PTC}) has an
additional term enabling the maintenance of PTC if no HH is expressed in the
neighboring cells. This term will not be modified by the two-step assumption,
only the dependence on the transcript, such that the new rule for PTC is
\begin{equation}
PTC^{t+1}_i=ptc_i^t \mbox { or } ptc_i^{t-1} \mbox { or } (PTC_i^t \mbox 
{ and not } HH_{i-1}^t \mbox{ and not } HH_{i+1}^t)
\end{equation}
The modification of CI into CIA and CIR is not directly affected  by the
two-step assumption, since it does not involve the {\it ci} transcript.
Nevertheless, it is no longer necessary to assume that the presence of the
{\it hh} transcript affects the outcome of the modification, since the two-step
maintenance of HH prevents the propagation of transient {\it hh} decays. Thus
the rules of creation of CIA and CIR simplify to
\begin{eqnarray}
CIA_i^{t+1}&=&CI_i^t \mbox { and } SMO_i^t\nonumber\\
CIR_i^{t+1}&=&CI_i^t \mbox { and not } SMO_i^t\nonumber
\end{eqnarray}
Finally, the rule for the production of the PH complex is unmodified by the
two-step assumption, as are the rules governing the transcription of {\it 
wg, en, hh, ptc and ci}.

Since we do not introduce new steps, but only delay certain steps in
the two-step model compared with the one-step model, the former leads
to exactly the same steady states as the latter. One way to explain
this result is that the system of equations to be solved in order to
find a fixed point of the Boolean rules is identical to
Equation \ref{eq_stable}, since in the fixed point
$x_i^{t+1}=x_i^t=x_i^{t-1}$ for the expression of any node. Another
way of illustrating the identity of the steady states in the two
models is by noting that the functional topology of the network
changes very little by the two-step assumption, and the {\it wg} and
PTC cycles are unmodified.

In addition we find that the two-step model reaches the same wild-type
steady state as that shown in Figure \ref{fig_wild}(b) if it is started
from the initial pattern of Figure \ref{fig_wild}(a).  The number of
intermediate states is slightly lower than in the one-step case, $5$
compared to $6$. These states arise mostly because the lack of
proteins in the initial state leads to the decay of the transcription
of several genes, and are not affected by the increased stability of
the proteins.  When the initial expression of a single gene is varied
while the others are wild type, the two-step model yields the same
steady states as the one-step model, with one notable exception: In
the case when {\it wg} is not expressed initially, the one-step model
leads to the state with no segmentation shown on
Figure \ref{fig_mutant}(b). However, the
two-step model settles into a periodic temporal repetition of $11$ states, only
five of them being devoid of an {\it engrailed/hedgehog - wingless}
boundary. This is the only case in which the final expression pattern
of the segment polarity genes is not time-invariant, and while this difference 
between the end
states is intriguing, it does not change the
conclusion that {\it wg} has to be prepatterned in order to reach
the wild-type steady state. Accordingly, the minimal prepatterning
leading to the wild-type final state remains the same as for the one-step model.

Our analysis shows that all the results presented in
Section \ref{sect_valid} are preserved if we use the two-step
assumption. Gene mutations and ectopic prepatternings lead to the same
steady states as in the original model.  The only change is in the
intermediate states visited en route to the final state: both the
wild-type and the broad type pattern stabilizes on average $30\%$
faster using the two-step assumption. On the other hand, the
pattern with no segmentation is reached at a slightly lower rate than in the original
model. In conclusion,  the two-step assumption provides a more
realistic modeling of the decay of the proteins without changing the
conclusions of the model.
 
\section{The expression of the segment polarity genes after a round of cell
division}
\label{later_stage}

When defining the Boolean functions characterizing the interactions
between nodes, we have assumed that the transport of WG and HH can be
disregarded. One of the reasons for doing so is the fact that we
assume that the parasegments are four cells wide, as they are at the
beginning of germ band elongation. Our results indicate that if we
allow the spreading of WG and HH further than the membrane of the nearest 
neighbors of the cells producing them,
ectopically broad $wg$, $en$, and $hh$ stripes result. This suggests
that during this stage the signaling of these secreted proteins is
limited to the interaction with the neighboring cells. This conclusion
is in agreement with the preference of the von Dassow model for low WG
transport rates (von Dassow {\it et al.} 2000, von Dassow \& Odell 2002). 
However, in later
stages the parasegment is enlarged due to two rounds of
divisions (Wolpert {\it et al.} 1998). While the $wg$ stripe remains a single cell
wide, the $en$ stripe widens to three cells. The maintaining of this
$en$ requires WG transport, and, indeed, wingless protein is seen to
diffuse over a distance of $2$-$3$ cell diameters in stage $10$
embryos (Bejsovec and Martinez-Arias 1991). It is possible that HH transport 
is also
stronger at this stage, although it is still limited by PTC.

\begin{figure}[htb!]
\centerline{\psfig{figure=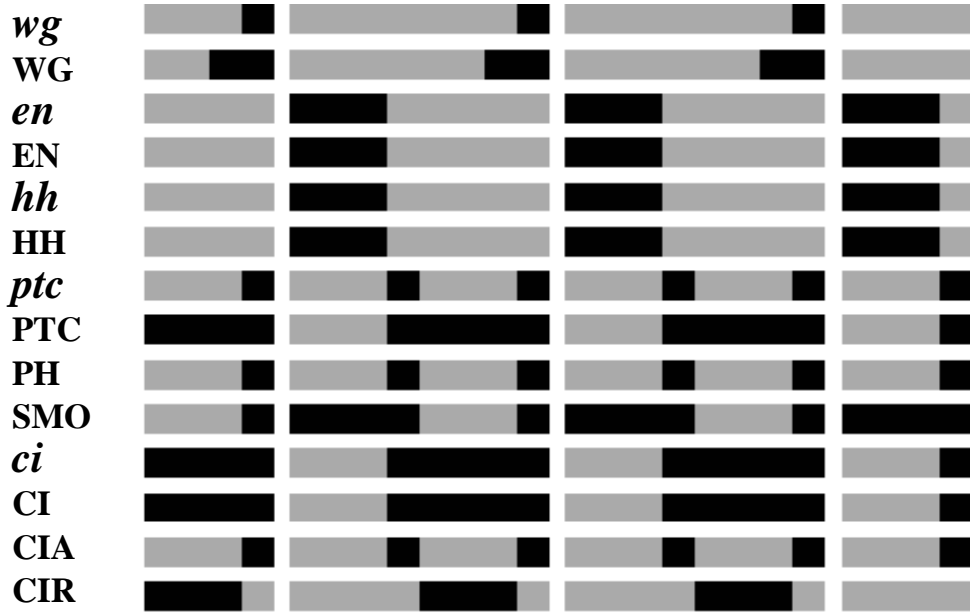,width=.5\textwidth,angle=-90}}
\caption{Stable expression pattern of the segment polarity genes after a round
of cell division, as obtained from our model. We assume that at 
this stage WG and HH can be transported through the neighboring cells. This
pattern is in good agreement with experimental observations of stage $11$
embryos.}
\label{fig_wide}
\end{figure}

In order to determine if our model is able to describe the segment polarity gene
patterns in later stages, we applied it to the transition between a four- and
eight-cell-wide parasegment. We started with the wild-type pattern of Figure
\ref{fig_wild}(a) and assumed that each cell divides into two identical cells,
with the same genes expressed in each of the two (see
Figure \ref{fig_wide}(a)).  We also assumed that WG and HH can be
transported through the nearest neighbors of the cells expressing their
mRNAs, and can interact with their receptors in the membrane of the second
neighbor cells. The model leads to the steady state represented in
Figure \ref{fig_wide}(b), with a single cell wide $wg$ stripe, three
cell wide $en$ and $hh$ stripes, and two $ptc$ stripes flanking the
$en$ domain. This steady state agrees perfectly with the wild-type
pattern observed in stage $11$ embryos (Hidalgo \& Ingham 1990, Eaton \& Kornberg
1990, Hooper \& Scott 1992, Alexandre {\it et al.} 1996, 
Alcedo {\it et al.} 2000). Note that in this state the WG stripe is broader
than the $wg$ stripe (as observed experimentally by Bejsovec and Wieschaus 
1993), but HH is expressed in the 
same cells as $hh$, its transport being restricted by the broad PTC domain.

\section{Discussion and outlook}

\begin{figure}[htb!]
\centerline{\psfig{figure=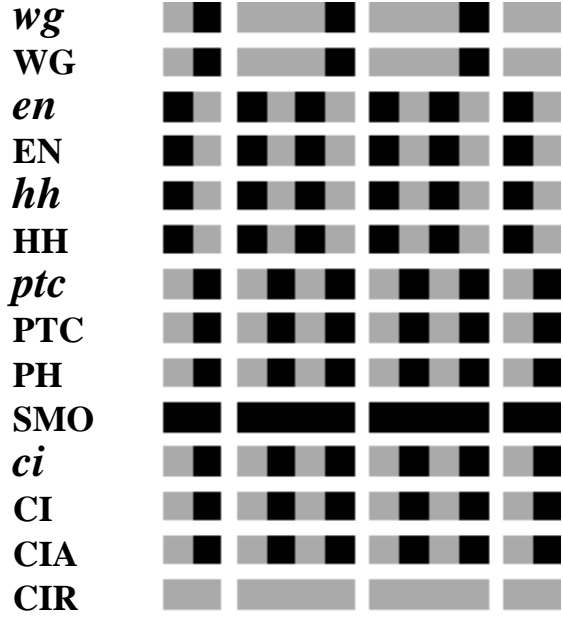,width=.5\textwidth,angle=-90}}
\caption{The pattern obtained from our model when we start from wild-type
initial conditions, but SLP is not functional. Note that $en$ is expressed in two stripes,
on both sides of the $wg$ stripe.}
\label{fig_nslop}
\end{figure}

Since this work focuses on the regulatory network of the same genes as in von
Dassow {\it et al.} (2000), it is worthwhile to discuss the common
points and differences between our results. Our model, based on the
topology and signature of the interactions between the segment polarity genes,
confirms their suggestion that the topology of the network has a dominant role
in its function. But it should be noted that the topology used in the von
Dassow {\it et al.} (2000) model and the present work is different. Since the
first reconstruction of von Dassow {\it et al.} did not capture the
asymmetrical activation of {\it en} by WG, they introduced two additional
interactions corresponding to intercellular WG transport and the inhibition of
{\it en} by CIR. While WG transport is relevant in the later stages of the
activity of the segment polarity genes (see Section \ref{later_stage}), the
inhibition of {\it en} was not observed experimentally. Instead, the
asymmetrical {\it en} activation is thought to be caused by the transcription
factors encoded by the {\it sloppy paired} gene (Cadigan {\it et al.} 1994), as
taken into account in our model. To test whether the activity of the SLP
proteins is necessary, we have studied the effects of inactivated and overexpressed
SLP. We obtain seven final states for inactivated SLP, but none of them
corresponds to the wild-type pattern. The closest state, obtained when we start
from wild-type initial conditions, has a wild-type $wg$ and $ptc$ pattern, but
$en$ and $hh$ have an ectopic stripe anterior to the $wg$ stripe, $ci$
separates into two thin stripes, and CIR is not expressed (Figure
\ref{fig_nslop}). At this point this state is a theoretical prediction that
can be verified by conditional SLP mutants ({\it i.e.}, mutants that have
normal pair-rule activity, but no segment polarity activity). We also find that
ubiquitously expressed SLP leads to the state with no segmentation presented on
Figure \ref{fig_mutant}(b). This finding is in agreement with the experimental
results of Cadigan {\it et al.} (1994). Based on these results we conclude that
the SLP proteins play a vital role in this network. 

Another difference between the von Dassow model and our model is the relative
importance of inhibitory and activating regulation. von Dassow {\it et al.}
consider inhibition as a modulation of the strength of activation (see the
equations in the Supplement to von Dassow {\it et al.} 2000), while we assume
that inhibition is dominant. The relatively weak role of inhibition in the
former probably
contributes to the large number of patterns with very broad {\it en} or {\it wg}
stripes found even for wild-type prepatterns, some of these patterns appearing much
more frequently (i.e. for a larger number of random parameter combinations) than
the wild-type pattern.

Our model is in very good agreement with known experimental results on the wild
type and mutant expression of the segment polarity genes. This gives numerous
insights into the internal robustness of the regulatory network, insights that
are important both from an experimental and theoretical perspective. First, we 
conclude that the {\it wingless} gene plays
a key role in the system, and it is imperative that it be initiated at the right
time in the right pattern. However, non-initiation of {\it engrailed} and
{\it hedgehog} can be rescued by the activity of the network. Experiments with 
conditional mutants defective in initiation could verify these predictions. 
Second, we find that the segment
polarity genes can evolve into an ectopic pattern with displaced stripes if 
initiated in a certain way.
While this ectopic initiation is difficult, it should be possible. Finally, the
isolation of the segment polarity- and pair-rule roles of the SLP proteins would
permit further insights into the functioning of the segment polarity gene
network.

Our goal was to remain as close to the experimentally-known facts as possible,
and to keep the number of additional assumptions and unknown parameters at a
minimum. In this spirit, we have assumed that the majority of the interactions
follow a single timescale. Two exceptions to this rule are the assumptions of 
the persistence of existent $wg$ and PTC expression in Eqs.
(\ref{eq_wg})  and (\ref{eq_PTC}). The stability of these nodes has a major 
role in stabilizing the
expression of the segment polarity genes. Indeed, these assumptions are
reflected in the existence of the cycles in the functional topology of
the network (Figure \ref{fig_topology}) and in the fact that the steady states are 
completely determined by the pattern
of $wg$ and PTC. It is therefore important to check what happens if these
assumptions are not used. If, instead of Equation (\ref{eq_wg}) we require that both
CIA and SLP be expressed in order that $wg$  be transcribed, {\it i.e.}, we assume
that
\begin{equation}
wg_i^{t+1}=CIA_i^t \mbox{ and } SLP_i^t \mbox{ and not } CIR_i^t,
\end{equation} 
it is still possible to arrive at the wild-type steady state, but only for much
more restricted initial states. Furthermore, the resilience of the network to
mutations in $ci$ is destroyed, as all initial states lead to the steady state
with no segmentation. Since it is observed experimentally that $ci$ null mutants
still display almost normal segmentation (Gallet {\it et al.} 2000), we can conclude that the
stability of $wg$ is required for the functioning of the segment polarity genes.
If we do not assume the maintenance of initial PTC expression, the pattern of
PTC will follow that of its transcript and split into two stripes. This will
cause the complete disappearance of CIR, normally expressed in the cells
in the middle of the PTC stripe, and the activation of all CI into CIA. As a
consequence, the wild-type steady state will be impossible to reach, instead,
the only steady state will be the pattern with broad stripes as in
Figure \ref{fig_mutant}(a). Thus both persistence assumptions are necessary, meaning
that the expression of $wg$ and PTC does not decay during the interval of
functioning of the segment polarity network.  

The two-step model represents a step towards modeling the
transition from the initial state to the final steady state of the
segment polarity network. A more realistic model would assume different time
intervals (expressed in number of steps) for the decay for mRNAs and
proteins.
While this extension would involve 
unknown parameters, the condition of reaching the same steady
states as the original model would provide constraints on the
variability of the decay rates. Another direction where our model
could be extended is to consider a two-dimensional array of cells. It
is known experimentally that the stripes of segment polarity genes are
not initiated as straight lines, but have jagged
borders (Wolpert {\it et al.} 1998). During the functioning of the segment polarity
network these stripes straighten, and the parasegment borders become
sharp. A two-dimensional simulation of our model could lead to
important insights into this process.  

While we have focused on the segment polarity genes here, the Boolean
approach can readily be applied to any gene regulation network with
relatively well-characterized interactions. The Boolean approach enables the 
integration of qualitative observations on gene interactions into a coherent picture,
and provides an easy verification of the sufficiency of these
interactions. The analysis of a Boolean model is more tractable than
that for a model based on differential equations, which inevitably has
numerous unknown parameters, and a Boolean model facilitates  a more
systematic study of the possible steady states and their
attractors. We envision realistic topology-based Boolean modeling as
an important first step in understanding the interplay between the
topology and functioning of gene regulatory networks. While the segment polarity
gene network was successfully modeled by a simple synchronous binary Boolean
model, other networks might require more detailed models
incorporating asynchronous updating and/or
multi-level variables (especially relevant for systems incorporating long-range
diffusion). Of course there are undoubtedly
systems, such as metabolic networks, for which a Boolean approach might not be an
appropriate first level of analysis.

\noindent {\bf Acknowledgement}

\noindent This work was supported in part by NIH Grant \#GM 29123.

\section{Appendix: Detailed rationale for our choice of Boolean updating rules}

The SLP proteins are translated from the {\it sloppy paired} ($slp$)
 gene that has both pair-rule and segment polarity roles
 (Grossniklaus 1992,  Cadigan {\it et al.} 1994). These authors show that 
 $slp$ is expressed in the posterior half of the parasegment primordium, and 
 this expression does not change
 during the developmental stages corresponding to the functioning of the segment
  polarity network. Since none of the nodes of the segment polarity
 network influence the translation of the SLP proteins (see
 Figure \ref{fig_gene}), we assume that their expression is maintained
 constant. 
\begin{equation}
SLP_i^{t+1}=SLP_i^t=\left\{\begin{array}{lllll}
0 &\mbox{if}& i\!\!\!\!\mod4=1 &\mbox{or}& i\!\!\!\!\mod4=2\\
1 &\mbox{if}& i\!\!\!\!\mod4=3 &\mbox{or}& i\!\!\!\!\mod4=0\\
\end{array}\right..
\label{eq_slp}
\end{equation}

The transcription of $wg$ is promoted by CIA (Aza-Blanc \& Kornberg 1999)
 and SLP (Cadigan {\it et al.} 1994), and repressed by CIR
(Aza-Blanc \& Kornberg 1999) (see Figure \ref{fig_gene}). According to
the assumptions listed above, this means that $wg_i^{t+1}=1$ in the
cells where $CIA^t_i=SLP^t_i=1$ and $CIR^t_i=0$. The transcription factors CIA
and SLP act independently (Cadigan {\it et al.} 1994, von Ohlen \& Hooper 1997),
thus we assume that the
effect of one of them is enough to maintain existing $wg$ expression ({ \it
i.e.}, the
transcriptional activation is stronger than the degradation of the mRNA).
Therefore, if $wg_i^t=1$, it is enough if either CIA$_i$ or
SLP$_i$ is present and CIR$_i$ is absent to obtain $wg_i^{t+1}=1$.
Consequently we have that 
\begin{equation}
wg_i^{t+1}=(CIA_i^t \mbox{ and } SLP_i^t \mbox{ and not } CIR_i^t) \mbox{ or } 
[wg_i^t \mbox{ and } (CIA_i^t \mbox{ or } SLP_i^t ) \mbox{ and not } 
CIR_i^t].
\label{eq_wg}
\end{equation}

WG is translated from $wg$, thus $WG_i^{t+1}=1$ if $wg_i^t=1$. WG is
secreted from the $wg$-expressing cells, and it is transported to
neighboring cells (Pfeiffer \& Vincent 1999, Hatini \& DiNardo 2001, von Dassow \& 
Odell 2002). The details
of the transport mechanism are not clear, but its
strength strongly depends on the developmental stage of the
embryo. In general, WG protein spreads from the $wg$-expressing cells
at a distance roughly equal to the width of the $en$ stripe
(Lecourtois {\it et al.} 2001). In the early stages of germ-band elongation WG is
 mainly associated with the cell membrane of the $wg$-expressing cells
(von Dassow \& Odell 2002), and mosaic embryo experiments done by  Vincent \&
Lawrence (1994) indicate that $en$
expression is sustained only in the cells adjoining $wg$.  In our  model the
presence of a substance implies the capability to influence others, and
consequently, we incorporate the lack of long-range WG signaling in this stage 
by assuming that WG is not transported further than the $wg$-expressing cells.
However, we take into account `degradation'
(which may in part be due to spatial spreading) by assuming that WG
expression is not maintained if $wg$ expression ceases, \ie\
$WG_i^{t+1}=0$ if $wg_i^t=0$. Thus the expression of WG depends only
of the expression of $wg$, and we have that 
\begin{equation}
\label{eq_WG}
WG_i^{t+1}=wg_i^t.
\end{equation}

The transcription of $en$ is promoted by WG in  neighboring cells
 (Cadigan \& Nusse 1997) and repressed by SLP (Cadigan {\it et al.} 1994). We assume
 that the presence of WG in either of the neighbor cells is enough to activate $en$
 transcription, thus
\begin{equation}
\label{eq_en}
en^{t+1}_i=(WG_{i-1}^t \mbox{ or } WG_{i+1}^t) \mbox{ and not } SLP^t_i.
\end{equation}

EN is translated from $en$, and therefore $EN^{t+1}_i=1$ if
$en^t_i=1$. Since EN is a transcription factor, it is assumed that
its expression will decay sufficiently rapidly that if $en^t_i=0$,
then $EN^{t+1}_i=0$, therefore
\begin{equation}
\label{eq_EN}
EN^{t+1}_i=en^t_i.
\end{equation}

$hh$ transcription is promoted by EN  (Tabata {\it et al.} 1992) and inhibited by
CIR (Aza-Blanc \& Kornberg 1999), and therefore
\begin{equation}
\label{eq_hh}
hh_i^{t+1}=EN_i^t \mbox { and not } CIR_i^t.
\end{equation}
HH is translated from $hh$, thus $HH_i^{t+1}=1$ if $hh_i^t=1$. HH is
secreted from the $hh$-expressing cells, and becomes bound to the cell
membrane, where it can bind to PTC present in one of the neighboring
cells, forming PH (Ingham \& McMahon 2001). It is also possible 
that HH is transported away from its source, as is the case in the wing
imaginal disks (Chen \& Struhl 1998, The {\it et al.} 1999, Ingham 2000). The proteoglycans involved
in HH transport compete with PTC in binding to HH, and thus PTC
expression restricts HH movement.  Since the PTC protein is expressed
everywhere except for the $en$/$hh$ expressing cells in this stage of
embryonic development, we do not take into account the signaling
effects of HH transport beyond the membrane of neighboring cells in this model.
In addition, we assume that
the HH levels decay if $hh$ is not expressed, due to  binding with 
PTC, degradation of the protein, etc.
\begin{equation}
\label{eq_HH}
HH^{t+1}_i=hh^t_i
\end{equation}

$ptc$ transcription is activated by CIA (Aza-Blanc \& Kornberg 1999) and 
repressed by CIR (Aza-Blanc \& Kornberg 1999).  It is likely that EN represses 
$ptc$ transcription (Hidalgo \& Ingham 1990), and thus $ptc$ is transcribed 
whenever CIA is present and CIR
and EN are absent:
\begin{equation}
ptc_i^{t+1}=CIA^t_i \mbox{ and  not } EN_i^t \mbox{ and not } CIR^t_i.
\end{equation}
Since PTC is translated from $ptc$, $PTC_i^{t+1}=1$ if $ptc_i^t=1$.
The HH protein from the neighboring cells binds to PTC, forming the PH
complex (Ingham \& McMahon 2001).  Since PTC is a
transmembrane receptor protein, and
the binding to HH is the only reaction PTC participates in, we assume
existing PTC levels are maintained even in the absence of $ptc$ if
there is no HH in the neighboring cells. 
\begin{equation}
PTC^{t+1}_i=ptc_i^t \mbox { or } (PTC_i^t \mbox { and not } HH_{i-1}^t 
\mbox{ and not } HH_{i+1}^t)
\label{eq_PTC}
\end{equation}

The PH complex forms when HH from  neighboring cells binds to
PTC. This binding is much faster than a transcription or translation
process, and we assume that it is instantaneous on the
chosen time scale. Since the PH complex may undergo endocytosis
(Taylor {\it et al.} 1993),  we assume that existing PH levels are not
maintained.
\begin{equation}
\label{eq_PH}
PH^{t}_i=PTC_i^t \mbox { and } (HH^t_{i-1} \mbox { or } HH^t_{i+1})
\end{equation}

{\it smo} is transcribed ubiquitously throughout the segment, but its
protein is deactivated post-translationally by PTC (Ingham 1998). 
Since we only consider the active substances in our model, we assume that
$SMO_i^t=0$ if $PTC_i^t=1$. Binding of HH from neighboring cells to
PTC relieves the inhibitory effect of PTC on SMO, leading to active
SMO. We assume that the conformational changes involved in the
deactivation and activation of SMO are instantaneous compared to the
time needed for transcription or translation, and thus $SMO_i^t=1$ if
$HH_{i\pm 1}^t=1$. The complete condition for activation of SMO is
therefore given as
\begin{equation}
SMO_i^t=\mbox{ not } PTC_i^t \mbox{ or } HH_{i-1}^t \mbox{ or } HH_{i+1}^t.
\label{eq_smo}
\end{equation} 
We do not explicitly relate the activation of SMO to the existence of
PH, as it is drawn on Figure \ref{fig_gene}, because the formation of PH
and the activation of SMO are the two outcomes of the same process
involving PTC$_i$, SMO$_i$ and HH$_{i\pm 1}$. While pictorially it is
more suggestive to draw a single edge from PH to SMO, as on
Figure \ref{fig_gene}, the basic condition for the activation of SMO is
the presence of HH$_{i\pm 1}$. Moreover, the PH-dependent logical
function, $SMO_i^t=\mbox{ not } PTC_i^t \mbox{ or } PH_i^t$ is
mathematically equivalent to (\ref{eq_smo}).

Since EN inhibits $ci$ transcription, $ci$ is expressed in the
complement of cells expressing EN (Eaton \& Kornberg 1990, 
Schwartz {\it et al.} 1995), and
therefore $ci^{t+1}_i=1$ if $EN^t_i=0$. Thus
\begin{equation}
ci_i^{t+1}=\mbox{not } EN_i^t. 
\end{equation}
CI is a cytoplasmic protein produced by $ci$  and
transformed into two nuclear proteins, CIA and CIR
(Aza-Blanc \& Kornberg 1999). Thus its expression decays if $ci$ is not 
expressed,
\ie\ $CI_i^{t+1}=0$ if $ci_i^t=0$. These two assumptions imply that CI
depends only on the expression of $ci$,
\begin{equation}
\label{eq_CI}
CI_i^{t+1}=ci_i^t.
\end{equation}

The CI protein is transformed to either CIA or CIR, depending on the
 state of SMO (Aza-Blanc \& Kornberg 1999, see also
Figure \ref{fig_gene}). If active SMO is present at time $t$, CI will be
transformed into CIA. Assuming that the
duration of this process is one timestep, the condition for
the existence of CIA at time $t+1$ is that $CI_i^t=1$ and $SMO_i^t=1$.
If all SMO is deactivated
by PTC, CI will be cleaved to form CIR; assuming that this
process takes one timestep, $CIR_i^{t+1}=1$ if $CI_i^t=1$ and $SMO_i^t=0$. Our
 results indicate
that these assumptions are not sufficient to reproduce the wild-type patterns of
CIA and CIR; instead, ectopic expression of CIR appears in
cells in which CIA should be expressed. The cause of this discrepancy
might be that the time needed for the post-translational modification of
CI is shorter than the duration of transcription and translation, and
$CIA_i^{t+1}$ is in fact influenced by SMO activated later than at
time $t$. One way to take this into account is to assume that the
presence of $hh$ at time $t$ in the neighboring cells may activate
SMO$_i$ and lead to the formation of CIA$_i$ at time $t+1$ if CI$_i$ is present
at time $t$. This is
incorporated as follows.
\begin{equation}
 CIA_i^{t+1}=CI_i^t \mbox { and } (SMO_i^t \mbox{ or } hh_{i-1}^t \mbox { or }
 hh_{i+1}^t).
\label{eq_CIA}
\end{equation}
CIR forms in the alternate pathway (see Figure \ref{fig_gene}), and
therefore relaxing the conditions for CIA formation will translate to a
restriction of CIR, as reflected in the rule 
\begin{equation}
 CIR_i^{t+1}=CI_i^t \mbox { and  not } SMO_i^t \mbox { and not }
 hh_{i\pm1}^t.  
\label{eq_CIR}
\end{equation}

\end{document}